\documentclass[pra,twocolumn,floatfix,nofootinbib]{revtex4-1}
\usepackage{amsmath,amssymb,amsthm}
\usepackage{graphicx} 
\usepackage{epstopdf}
\usepackage{bbm}
\usepackage[colorlinks,allcolors=blue]{hyperref}
\usepackage[up]{subfigure}
\usepackage{commath}

\makeatother

\begin{document}

%\preprint{APS/123-QED}

\title{Coherence and Entanglement Monogamy in \\ the Discrete Analogue of Analog Grover Search}% Force line breaks with \\
%\thanks{Pun intended!}%

\author{Namit Anand}
\email{namit.anand@niser.ac.in}
\affiliation{Department of Physics\\
National Institute of Science Education and Research, Bhubaneswar,
752050, India}
% \altaffiliation[Also at ]{Independent}%Lines break automatically or can be forced with \\
\author{Arun Kumar Pati}%
 \email{akpati@hri.res.in}
\affiliation{%
 Quantum Information and Computation Group\\
Harish-Chandra Research Institute, Chhatnag Road, Jhunsi, Allahabad 211019, India
}%

\date{\today}% It is always \today, today,

\begin{abstract}
Grover's search algorithm is the optimal quantum algorithm that can search an unstructured database quadratically faster than any known classical algorithm. The role of entanglement and correlations in the search algorithm have been studied in great detail and it is known that entanglement between the qubits is necessary to gain a quadratic speedup, for pure state implementation of the Grover search algorithm. Here, we systematically investigate the behavior of quantum coherence and monogamy of entanglement in the discrete analogue of the \textit{analog analogue of Grover search algorithm}. The analog analogue of Grover search is a continuous time quantum algorithm based on the adiabatic Hamiltonian evolution that gives a quadratic speedup, similar to the original Grover search algorithm. We show that the decrease of quantum coherence, quantified using various coherence monotones, is a clear signature of attaining the maximum success probability in the analog Grover search. We also show that for any two qubit reduced density matrix of the system, the concurrence evolves in close vicinity to the increasing rate of success probability. Furthermore, we show that the system satisfies a $n$-party monogamy inequality for arbitrary times, hence bounding the amount of $n$-qubit entanglement during the quantum search.

%\begin{description}
%\end{description}
\end{abstract}

\pacs{Valid PACS appear here}% PACS, the Physics and Astronomy
\maketitle

\section{Introduction}
\label{sec:introduction}
The idea that quantum mechanical systems can efficiently simulate physical systems \cite{ref1} is at the heart of the theory of quantum information and computation. Deutsch \cite{ref2} constructed the first example of an algorithm which would require two queries to solve on a classical computer but that can be solved with only one quantum query. Subsequently, Deutsch and Jozsa \cite{ref3}, Bernstein and Vazirani \cite{ref4} and Simon \cite{ref5} demonstrated the striking difference between the classical and quantum query complexity. This culminated with Grover's search algorithm \cite{ref6} achieving a quadratic speedup over any classical algorithm for the unstructured search problem along with Shor's factoring algorithm \cite{ref7} that could factor integers efficiently and also calculate discrete logarithms.

Grover's algorithm gives an optimal solution to the unstructured search problem, the problem of deciding whether a black-box Boolean function has any input that evaluates to 1. It provides a quadratic temporal speedup over the best classical search algorithms, even when they both require the same spatial resources to perform the search task. Zalka \cite{ref8} proved the optimality of the Grover search algorithm.  Grover's algorithm was subsequently generalized to the framework of amplitude amplification and to counting the number of solutions by Brassard \textit{et al.} \cite{ref9}. Grover's search algorithm has since been applied to a wide variety of similar problems \cite{ref10}\cite{ref11}\cite{ref12}.

The usual paradigm of computation (quantum or classical) is defined in a discrete setting. However, adiabatic quantum computation \cite{ref13} provides a continuous time model for quantum computing by using the quantum adiabatic theorem. Here, one begins by finding an appropriate Hamiltonian whose ground state describes the solution to the problem of interest. Next, a system with a simple Hamiltonian is prepared and initialized to the ground state. Finally, the simple Hamiltonian is adiabatically evolved to the desired Hamiltonian. By the quantum adiabatic theorem, the system remains in the ground state at all times during this adiabatic evolution, depending on the spectral properties of the Hamiltonian \cite{ref14}\cite{ref15}. At the end, the final state of the system describes the solution to the problem. The time complexity for an adiabatic algorithm is simply defined as the time taken to complete the adiabatic evolution, which is known to depend on the spectral gap of the Hamiltonian \cite{ref15}. The adiabatic model of quantum computing is universal and equivalent to standard quantum computation \cite{ref16}. Farhi and Gutmann were the first to construct an analog algorithm for solving the Grover search problem \cite{ref17} referred to hereafter as the \textit{analog analogue of Grover search algorithm} or simply the \textit{analog Grover search algorithm}. 

It is understood that the speedup in quantum algorithms comes from quantum mechanical features like quantum coherence and quantum correlations like quantum entanglement, that exist amongst the qubits. Quantum correlations, especially quantum entanglement, is one of the crucial resources in quantum information theory and has been studied extensively \cite{ref18}. The indispensable role of entanglement in quantum information processing tasks such as quantum teleportation \cite{ref19}, superdense coding \cite{ref20} and remote state preparation \cite{ref21new} \cite{ref21} , etc. has been quantified and understood deeply. However, despite showing that quantum entanglement is \textit{necessary} for the pure state implementation of Grover search \cite{ref22} \cite{ref22new} and Shor's factoring algorithm \cite{ref23}, the innate role of entanglement is not clear for general quantum computing tasks. 

One of the fundamental attributes of quantum systems is their ability to exist in linear superpositions of different physical states. This physical phenomenon is called quantum superposition. Quantum coherence arises from superposition and is at the heart of several quantum features like quantum interference, multiparticle entanglement \cite{ref24}, quantum biology  \cite{ref25} \cite{ref25new} \cite{ref26}, quantum thermodynamics \cite{ref27} \cite{ref28} \cite{ref29} \cite{ref30}, quantum game theory  \cite{ref31} \cite{ref32} \cite{ref33} etc., which in turn are some of the most important applications of quantum physics and quantum information science. Inspired from the resource theory of entanglement \cite{ref18} \cite{ref35}, there has been a lot of effort recently to quantify coherence as a resource theory \cite{ref36}. The role of coherence in the Deutsch-Jozsa algorithm has also been explored recently \cite{ref36new}. In a standard resource-theoretic treatment of coherence, the ``incoherent'' states are those that are diagonal in some fixed reference basis. The amount of coherence is then defined as the distance from these reference states. A `coherence measure' is a real-valued function over the quantum state-space, such that it vanishes for all the states that are deemed to be incoherent and cannot increase under some class of operations that preserve incoherence. 

In this paper, we quantitatively analyze the role of quantum coherence and monogamy of entanglement \cite{ref37} \cite{ref38} in the discrete analogue of the analog Grover search algorithm. To quantify the dynamics of coherence in this analog Grover search, we use two coherence monotones namely the $l_1$-norm and the relative entropy of coherence \cite{ref36}. We then discretize the analog Grover search algorithm to study how the monogamy of quantum entanglement evolves parallel to the system's adiabatic evolution. We use an entanglement monotone called the concurrence~\cite{ref39} to study the same. 

We find that the amount of coherence (as quantified by the coherence measures) is non-zero at all times during the search. The only time this coherence goes to zero is when the search algorithm attains the maximum success probability equal to one. Since we begin with a maximally coherent state, in a resource theoretic sense, $n$-maximal coherence is actually consumed during the quantum search. Using coherence monotones, we relate the success probability of the search algorithm with the amount of coherence. Further, during the analog search, the final state is a product state and so all entanglement monotones must go to zero at the end of the search algorithm. The bipartite entanglement entropy, concurrence and the monogamy score, all behave in accord with this, since for both qubit partitions, these measures of quantum correlations go to zero and mark the completion of the search algorithm. Furthermore, the two qubit concurrence and the monogamy score peak simultaneously with the increasing rate of success probability, hence implying that entanglement monogamy is indeed satisfied for the discrete analogue of the analog analogue of Grover search algorithm. 

The paper is organized as follows. In Sec. \ref{sec:review}, we review the analog Grover search algorithm. In Sec. \ref{sec:coherence}, we use the two coherence monotones to calculate the dynamics of coherence and elucidate their behavior. We also define the mapping through which we discretize the analog Grover search algorithm. In Sec. \ref{sec:entanglement}, we explore the entanglement entropy and the concurrence, and demonstrate the synonymous behavior between the rate of change of concurrence in the continuous and the discrete Grover search. In the next section, we calculate the two-qubit concurrence and show its connection to the success probability of the search algorithm. We then use the monogamy inequality to bound the bipartite entanglement at all times during the search. We conclude our paper with the discussions in Sec. \ref{sec:conclusion}.

\section{Analog Grover search algorithm}
\label{sec:review}
Adiabatic quantum computation can be described as a controlled Hamiltonian evolution of a system obeying the Schr\"{o}dinger equation
\begin{equation}
i \frac{d}{dt} |\psi \rangle = H(t) |\psi \rangle.
\end{equation}
We briefly recapitulate the analog analogue of Grover search algorithm~\cite{ref17}, where the problem is to use quantum evolution to find a marked state among $N$ orthonormal states.\\

Imagine that we are given a Hamiltonian in an $N$ dimensional Hilbert space such that it has only one non-zero eigenvalue, $E\neq 0$.  The task then is to find the corresponding eigenvector $| w\rangle$ which has eigenvalue $E$. The Hamiltonian can be represented as
\begin{equation}
H_w = E | w\rangle \langle w|
\end{equation}
with $| w\rangle$ unknown and normalized, i.e., $\langle w|w\rangle=1$.  Now we choose some normalized vector $|s\rangle$, which is independent of $| w\rangle$ (since we do not know what $| w\rangle$ is yet). We then add to $H_w$, the ``driving" Hamiltonian \cite{ref17}
\begin{equation}
H_D = E|s\rangle \langle s|
\end{equation}
so that the full Hamiltonian is given by
\begin{equation}
H=H_w + H_D.
\end{equation}
Now, starting from the initial state $|s\rangle$ at $t=0$, we calculate the time evolution of the state $|\psi (t) \rangle$. Since the total Hamiltonian is time-independent, we can simply write the time evolved state as
\begin{equation}
|\psi(t)\rangle = e^{-iHt} \ |s\rangle.
\end{equation}
We work in the units $\hbar = 1$. It suffices to confine our attention to the two dimensional subspace spanned by
$|s\rangle$ and $|w\rangle$.  The vectors $|s\rangle$ and $|w\rangle$ are not orthogonal (in general) and let us denote their inner product as $x$, i.e., $\langle s|w\rangle = x $, where $x$ can be taken to be real and positive since any phase in the inner product $\langle s|w\rangle$ can be absorbed in  $|s\rangle$. Now the vectors
\begin{equation}
|r\rangle = \frac{1}{\sqrt{1-x^2}} \left( |s\rangle - x | w\rangle\right)
\end{equation}
and $|w\rangle$ are orthonormal.  In the \{$|w\rangle$, $| r\rangle$\} basis, the
Hamiltonian is given by
\begin{equation}
H = E \left[\begin{array}{cc} 1 + x^2 & x\sqrt{1-x^2} \\ x\sqrt{1-x^2} & 1-x^2
\end{array}\right]
\end{equation}
and
\begin{equation}
|s\rangle = \left[\begin{array}{c} x \\ \sqrt{1-x^2} \end{array}\right].
\end{equation}
Now the state of the system at time $t$ is found to be
\begin{equation}
\label{eqn:wavefunction}
|\psi(t)\rangle = e^{-iEt} \left[\begin{array}{c} x \ \cos (E  x  t)  - i
\ \sin (E  x t)
\\
\sqrt{1-x^2}  \cos (E  x  t) \end{array}\right] .
\end{equation}
Thus, we can see that at time $t$, the probability of finding the state $|w\rangle$ is given by
\begin{equation}
P(t)=\sin^2  (E  x  t) + x^2 \ \cos^2  (E  x  t)
\label{eqn:successprobability}
\end{equation}
and the probability is one at time $t_m$ given by
\begin{equation}
t_m = \frac{\pi}{2Ex} .
\label{eqn:time}
\end{equation}

The inner product between the vectors  $|w\rangle$ and  $|s\rangle$ (defined as \emph{x} above), is assumed to be non-zero, else it will take infinite time for the quantum search to complete.

This analog model has been generalized further. For example, the same algorithm was recast in the form of a spatial lattice search problem \cite{ref40}, where there is an $N$-dimensional lattice and the basis state $|i \rangle$ is localized at the $i$th lattice site. The on-site potential energy, $E$ is zero everywhere except at $|w \rangle$, where it takes the value 1. The objective is same as before, to reach the marked state $|w \rangle$ starting from an equal superposition of all the $|i\rangle$ 's. The kinetic term is formulated through the Laplacian of the lattice, which effectively introduces uniform hopping to all the nearest neighbors from any given lattice site, and is kept constant. Our results will also hold in these generalized models.

\section{Quantum coherence in Grover search}
\label{sec:coherence}
Quantum coherence is one of the salient features of the quantum world. As this crucial feature drives several quantum technologies, it is very desirable to quantify the usefulness of coherence as a resource. This is done using the mathematical framework called a `resource theory'. To characterize something as a resource, we must first impose certain constraints on what specific physical operations are allowed (e.g. the local operations and classical communication(LOCC) framework for quantum entanglement \cite{ref43} restricts one from performing joint quantum operations between distant laboratories), which define the freely accessible operations. To be able to execute general quantum operations under such a constraint then requires some ``special'' quantum states that contain a relevant resource (e.g. entangled states) and can be consumed in the process. In fact, a quantitative resource theory for entanglement is already in place \cite{ref18} \cite{ref35} and has later expanded to encompass a wider range of operational phenomenon \cite{ref44} \cite{ref45}.

However, for a long time there did not exist a resource theoretic framework to quantify the physical aspects of coherence. Coherence was often defined in terms of functions of the off-diagonal entries of a density matrix, and the definition justified on the basis of physical intuition, which in turn also lead to the idea of decoherence developing along similar lines \cite{ref46}. Eventually, in 2014, Baumgratz \textit{et al}. \cite{ref36} defined a resource theory of coherence and rigorously quantified the role of coherence in close analogy to the resource theory of entanglement.

Let us formally introduce the measures of coherence in the framework of a resource theory that is based on the set of incoherent operations and incoherent states \cite{ref36}. Coherence is a basis dependent property and hence first, we must fix a  \textit{reference basis}. The choice of the reference basis may be dictated by the underlying physics of the problem (say the energy eigenbasis) or the task to be performed (wherever we wish to use quantum coherence). Let \{$|i \rangle_{i=0,1,2,...,d-1}$\} be a basis for $\mathcal{H}^d$, a $d$-dimensional Hilbert space. The set of all density matrices that are diaogonal in this chosen basis form the so called ``incoherent states''. The set of all the states of the form $\rho_I= \sum_{i}p_i |i\rangle \langle i|$, where $p_i\geq0$ and $\sum_i p_i=1$, forms the set $\mathcal{I}$ of all incoherent states. Any quantum state that does not belong to the set $\mathcal{I}$ will be called a coherent state and will act as a resource. A maximally coherent state in this basis is then given by $|\psi_d \rangle = \frac{1}{\sqrt{d}}\sum_{i=0}^{d-1} |i \rangle$, as any other state can be created from $|\psi_d \rangle$ using only the set of `incoherent operations'. Baumgratz \textit{et al}. \cite{ref36} define incoherent operations as the set of trace preserving completely positive maps $\Lambda: B(H) \longmapsto B(H)$ admitting a set of Kraus operators ${K_n}$ \cite{ref41} such that $\sum_n {K_n}^{\dagger} K_n = \mathbbm{1}$ (trace preservation) and, for all $n$ and $\rho \in \mathcal{I}$,
\begin{equation}
\frac{K_n \rho {K_n}^{\dagger}}{Tr[K_n \rho {K_n}^{\dagger}]} \in \mathcal{I}.
\end{equation}
The definition of the set of incoherent operations is not unique and so we redirect the reader to a recent review for further discussions \cite{ref47}. We shall use two coherence monotones introduced in \cite{ref36}, namely, the $l_1$-norm of coherence and the relative entropy of coherence.

Given a density matrix $\rho$ for the system, the $l_1$-norm of coherence is defined as
\begin{equation}
\mathcal{C}_{l_1} (\rho) = \sum_{i,j; i\neq j} \abs{\rho_{i,j}}.
\end{equation} 

For the analog Grover search, the $l_1$-norm of coherence for the state given in Eq. (\ref{eqn:wavefunction}) is found to be
\begin{small}
\begin{equation*}
\mathcal{C}_{l_1} (\rho) = 2 \abs{\cos (E x t)} \sqrt{\left(1-x^2\right) \left(x^2 \cos ^2(E x t)+\sin ^2(E x t)\right)}.
\end{equation*}
\end{small}

%hiding this
%\begin{figure}[hbtp]
%\centering
%\includegraphics[scale=0.7]{coherence1newcolor.png}
%\caption{$l_1$ norm of coherence and probability of success vs time at $E=1$, $x$ = 0.707.}
%\end{figure}

%\begin{figure}[hbtp]
%\centering
%\includegraphics[scale=0.45]{coherence1.png}
%\caption{Variations of the $l_1$-norm of coherence and the probability of success as a function of time at parameter values $E=1$ and $x$ = 0.707.}
%\label{fig:l1norm}
%\end{figure}
%
Similarly, the relative entropy of coherence is given by
\begin{equation}
\mathcal{C}_{r} (\rho) = S(\rho_{diag}) - S(\rho),
\end{equation} 
which is found to be
\begin{small}
\begin{flalign*}% left aligned
&\mathcal{C}_{r}  (\rho) = - \left(1 - x^2\right) \cos ^2(E x t) \log_2 \left(\left(1- x^2\right) \cos ^2(E x t)\right) &\\ 
&-\left(x^2 \cos ^2(E x t)+\sin ^2(E x t)\right) \log_2 \left(x^2 \cos ^2(E x t)+\sin ^2(E x t)\right).&
\end{flalign*}
\end{small}
%hiding this
%\begin{figure}[hbtp]
%\centering
%\includegraphics[scale=0.65]{coherence2newcolor.png}
%\caption{Relative entropy of coherence and probability of success vs time at $E=1$, $x$ = 0.707.}
%\end{figure}

%\begin{figure}[hbtp]
%\centering
%\includegraphics[scale=0.45]{coherence2.png}
%\caption{Variations of the relative entropy of coherence and the probability of success as a function of time at parameter values $E=1$ and $x$ = 0.707.}
%\label{fig:relcoherence}
%\end{figure}

\begin{figure}[hbtp]
\centering
\includegraphics[scale=0.5]{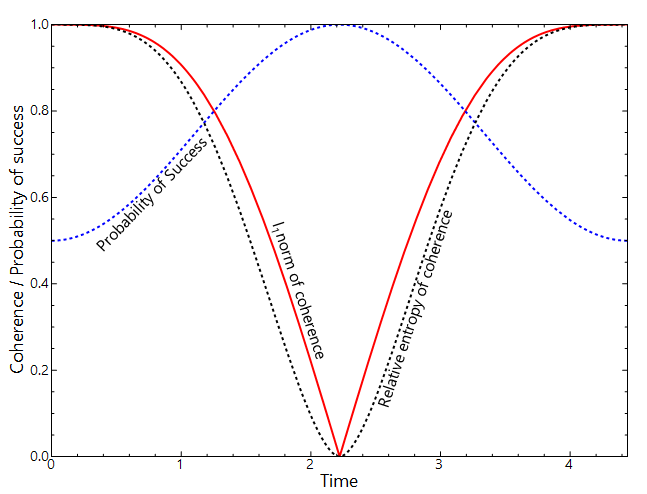}
\caption{Variations of the $l_1$-norm, the relative entropy of coherence and the probability of success as a function of time at parameter values $E=1$ and $x$ = 0.707.}
\label{fig:relcoherence}
\end{figure}

The $l_1$-norm of coherence has a closed form expression in terms of the probability of success given by, 
\begin{equation}
\label{eqn:l1intermsofp}
\mathcal{C}_{l_1} (P) = 2 \sqrt{P(1-P)},
\end{equation}
where the form of $P$ is given in Eq. (\ref{eqn:successprobability}). Similarly, the relative entropy of coherence expressed in terms of the probability of success takes the form of a binary entropy function, i.e., 
\begin{equation}
\label{eqn:relintermsofp}
\mathcal{C}_{r} (P) = - \left(P \log_2{P} + (1-P) \log_2{(1-P)} \right).
\end{equation}

It can be observed analytically from Eq. (\ref{eqn:l1intermsofp}) and Eq. (\ref{eqn:relintermsofp}) and numerically from the Fig. \ref{fig:relcoherence} that both coherence measures go to zero \textit{iff} the probability of success peaks to one, hence acting as a clear signature for the success of the search algorithm. Even for other general initial states (i.e., with a different value of $x$), after evolving under the analog Grover search Hamiltonian, the coherence attains a minima as the success probability peaks to 1. Note that although $P=0$ would also imply both coherence measures going to zero as can be observed from Eq. (\ref{eqn:l1intermsofp}) and Eq. (\ref{eqn:relintermsofp}), the probability of success cannot be zero since that would require $x$ to be zero (which is excluded else the search would take infinite time to complete, see Eq. (\ref{eqn:time})).

\subsection{Coherence in discrete analogue of analog Grover search algorithm}
\label{sec:mapping}

To quantify coherence in the analog Grover search algorithm, we consider a mapping through which the $N$ eigenstates of the system are mapped to the logical states of $n$ qubits (where $N = 2^{n}$). The Hamiltonian initially acting on the state $|\psi \rangle$ of the system now acts on the $n$ qubit system. Throughout this paper, we refer to this mapping between the $N$ eigenstates and the $n$ qubits  as the \textit{discrete analogue of analog Grover search algorithm}.\\ 

Now, consider the initial state to be an equal superposition of all the basis states, i.e.,\\
\begin{equation}
|s \rangle = \frac{1}{\sqrt{N}}\sum_{i=0}^{N-1} |i\rangle
\end{equation}
and the final state $|w\rangle$ is one of the orthonormal basis states. The $|\psi(t)\rangle$ can be expressed in the \{$|w\rangle,|r\rangle$\} basis as
\begin{equation}
|\psi(t)\rangle = \alpha(t)|w\rangle + \beta(t)|r\rangle,
\end{equation}
where $\alpha=x\cos(Ext)-i\sin(Ext)$, $\beta=\sqrt{1-x^2}\cos(Ext)$ and $x=\frac{1}{\sqrt{N}}$.\\

Therefore, the density matrix in the \{$|w\rangle,|r\rangle$\} basis is
\begin{equation}
\label{eqn:densitymatrix}
\begin{split}
\rho = \alpha^2(t)|w \rangle \langle w| + \alpha(t)\beta(t)|w \rangle \langle r| \\ + \alpha^*(t)\beta(t)|r \rangle \langle w| +  \beta^2(t)|r \rangle \langle r|.
\end{split}
\end{equation}

After discretizing the analog Grover search, the $l_1$-norm of coherence for the density matrix in Eq. (\ref{eqn:densitymatrix}) is analytically calculated to be
\begin{small}
%\begin{flalign*}% left aligned
%\mathcal{C}_{l_1} (\rho) = & 2 \sqrt{\left(1-\frac{1}{N} \right) \left(\frac{1}{N} \cos ^2(\frac{E t}{\sqrt{N}})+ \sin ^2(\frac{E t}{\sqrt{N}}) \right)} &\\ & \text{  x  } \abs{\cos (\frac{E t}{\sqrt{N}})}. &
%\end{flalign*}
%
\begin{equation*}
\mathcal{C}_{l_1} (\rho) = 2 \abs{\cos (\frac{E t}{\sqrt{N}})} \sqrt{\left(1-\frac{1}{N} \right) \left(\frac{1}{N} \cos ^2(\frac{E t}{\sqrt{N}})+ \sin ^2(\frac{E t}{\sqrt{N}}) \right)}.
\end{equation*}
\end{small}

Similarly the relative entropy of coherence is calculated to be
\begin{widetext}
\begin{small}
\begin{equation*}
\mathcal{C}_{r}  (\rho) = - \left(1 -\frac{1}{N} \right) \cos ^2(\frac{E t}{\sqrt{N}}) \log_2 \left(\left(1- \frac{1}{N}\right) \cos ^2(\frac{E t}{\sqrt{N}})\right) 
-\left(\frac{1}{N} \cos ^2(\frac{E t}{\sqrt{N}})+\sin ^2(\frac{E t}{\sqrt{N}})\right) \log_2 \left(\frac{1}{N} \cos ^2(\frac{E t}{\sqrt{N}})+\sin ^2(\frac{E t}{\sqrt{N}})\right).
\end{equation*}
\end{small}
\end{widetext}

%\begin{small}
%\begin{flalign*}% left aligned
%&\mathcal{C}_{r}  (\rho) = - \left(1 -\frac{1}{N} \right) \cos ^2(\frac{E t}{\sqrt{N}}) \log_2 \left(\left(1- \frac{1}{N}\right) \cos ^2(\frac{E t}{\sqrt{N}})\right) &\\ 
%&-\left(\frac{1}{N} \cos ^2(\frac{E t}{\sqrt{N}})+\sin ^2(\frac{E t}{\sqrt{N}})\right) \log_2 \left(\frac{1}{N} \cos ^2(\frac{E t}{\sqrt{N}})+\sin ^2(\frac{E t}{\sqrt{N}})\right).&
%\end{flalign*}
%\end{small}

The $l_1$-norm and the relative entropy of coherence display a similar behavior as before, and it is important to note that the initial state for the analog Grover search algorithm is an equal superposition of all orthonormal basis states, i.e.,
\begin{equation}
|s \rangle = \frac{1}{\sqrt{N}}\sum_{i=0}^{n-1} |i\rangle ,
\end{equation}
which corresponds to a maximally coherent state in this eigenbasis. However, at time $t = \frac{\pi \sqrt{N}}{2E}$ (which is the time at which the search algorithm succeeds), the coherence is reduced to zero. This implies that maximally coherent states are actually consumed during the search algorithm.

\section{Entanglement monogamy in Grover search}
\label{sec:entanglement}
\subsection{Entanglement entropy}
Entanglement entropy is a measure of entanglement for many body quantum states. Bipartite entanglement entropy is defined with respect to a bipartition of a state into two parts say $A$ and $B$. For this, consider a quantum system whose Hilbert space $\mathcal{H} = \mathcal{H}_A \otimes \mathcal{H}_B$. The bipartite entanglement entropy $S$ of a state $|\Psi \rangle \in \mathcal{H}$ is defined as the von Neumann entropy of either of its reduced states. That is, for a pure state $ \rho_{AB}= |\Psi\rangle\langle\Psi|_{AB} $, the entanglement entropy is given by\\

$\mathcal{S}(\rho_A)=  -\operatorname{Tr}(\rho_A\operatorname{log}\rho_A) =  -\operatorname{Tr}(\rho_B\operatorname{log}\rho_B) = \mathcal{S}(\rho_B),$
\\ \\where $ \rho_{A}=\operatorname{Tr}_B(\rho_{AB}) $ and $ \rho_{B}=\operatorname{Tr}_A(\rho_{AB}) $ are the reduced density matrices for each partition. The von Neumann entanglement entropy in the eigenbasis is defined as
\begin{equation}
S(\rho) = -\sum^{n}_{i=0} \lambda_i \log{\lambda_i}.
\end{equation}

It is easier to quantify the role of entanglement in the original Grover search due to its discrete nature, however it is not so in the analog version. Therefore, we use the discrete mapping to calculate the entanglement entropy for the analog Grover search. We partition the $n$ qubit state into one qubit (say $l$) vs  the other $n-1$ qubits. The reduced density matrix for the $l$th qubit, $\rho_l$ is obtained by tracing out the other $(n-1)$ qubits and is given as
\begin{large}
\begin{equation}
\rho_l = \begin{bmatrix}{}
 \abs{\alpha}{}^2 + \frac{(N-2) \beta^2}{2 N-2}  & \frac{(N-2) \beta^2}{2 N-2}+\frac{\alpha \beta}{\sqrt{N-1}} \\
 \frac{(N-2) \beta^2}{2 N-2}+\frac{\alpha^* \beta}{\sqrt{N-1}} & \frac{\beta^2 N}{2 N-2} \\
\end{bmatrix} .
\label{eqn:singlequbit}
\end{equation}
\end{large}

The eigenvalues of the matrix, $\lambda_{\pm}$ are given by
\begin{equation}
\lambda_{\pm} = \frac{2 N \pm \sqrt{N \left((N-2) \cos \left(\frac{4 E t}{\sqrt{N}}\right)+3 N+2\right)}}{4 N} ,
\label{eqn:eigenvalues}
\end{equation}
from which one can find the $S(\rho_l)$.

%Using the eigenvalues listed above, the entanglement entropy for the system partition of $1:(n-1)$ is given by
%\begin{small}
%\begin{equation*}
%S = \begin{split}
%- \left(\frac{2 N-\sqrt{N \left((N-2) \cos \left(\frac{4 E t}{\sqrt{N}}\right)+3 N+2\right)}}{4 N}\right)
%\\
% \log_2 \left(\frac{2 N-\sqrt{N \left((N-2) \cos \left(\frac{4 E t}{\sqrt{N}}\right)+3 N+2\right)}}{4 N}\right)
%\\
% - \left( \frac{2 N+\sqrt{N \left((N-2) \cos \left(\frac{4 E t}{\sqrt{N}}\right)+3 N+2\right)}}{4 N} \right)
%\\ 
% \log_2 \left( \frac{2 N+\sqrt{N \left((N-2) \cos \left(\frac{4 E t}{\sqrt{N}}\right)+3 N+2\right)}}{4 N} \right)
%\end{split}.
%\end{equation*}
%\end{small}
%hiding this
%\begin{figure}[hbtp]
%\centering
%\includegraphics[scale=0.65]{3newnewcolor.png}
%\caption{Entanglement entropy and probability of success vs time at $E=1$, $N$ = 4.}
%\end{figure}
%
%\begin{figure}[hbtp]
%\centering
%\includegraphics[scale=0.45]{entanglemententropy.png}
%\caption{Variations of the entanglement entropy(across system partition of $A_1 | A_2 A_3 ... A_N $) and the probability of success as a function of time at parameter values $E=1$ and $N=4$.}
%\label{fig:entropy}
%\end{figure}

%\begin{figure}[hbtp]
%\centering
%\includegraphics[scale=0.7]{4.png}
%\caption{Rate of change of entanglement entropy (i.e. $\frac{ds}{dt}$)vs time at $n$ = 4.}
%\end{figure}
%The evolution of Entanglement entropy is in direct correlation to that of the discrete Grover search algorithm.

\subsection{Concurrence}
\label{subsec:concurrence}
Concurrence is an entanglement measure, which for the special case of a pair of qubits is closely related to the entanglement of formation \cite{ref39}. For an arbitrary two qubit density matrix $\rho$, concurrence is defined as follows: we first define a ``spin-flipped'' density matrix, $\gamma$ as $(\sigma_y \otimes \sigma_y) \rho^* (\sigma_y \otimes \sigma_y) $ where $\sigma_y$ is the Pauli matrix $\left(
\begin{array}{cccc}
0 &-i\\ i &0
\end{array}\right)$. Then we calculate the square root of the eigenvalues of the matrix ($\rho \gamma$) and arrange them in decreasing order (say $\lambda_1$,$\lambda_2$,$\lambda_3$,$\lambda_4$). The concurrence is then defined as 
\begin{equation}
C(\rho) = \text{max } (\lambda_1-\lambda_2-\lambda_3-\lambda_4,0).
\end{equation}
For pure states, the concurrence is also defined via another quantity known as the ``tangle'', where 
\begin{equation}
\tau (|\psi \rangle) = 2(1-Tr(\rho^2))
\end{equation}
and the $C(\rho)$ = $\sqrt{\tau}$.

Concurrence for the pure state (in Eq.(\ref{eqn:wavefunction})) across the bipartition $A_1 | A_2 A_3 ... A_N $ is found to be
%\begin{large}
\begin{equation}
C(\rho_{A_1| A_2 A_3 ... A_N }) = \sqrt{\frac{(N-2)}{2N}} \abs{\sin{\left(\frac{2 E t}{\sqrt{N}}\right)}}.
\end{equation}
%\end{large}
%hiding this
%\begin{figure}[hbtp]
%\centering
%\includegraphics[scale=0.65]{conc1newcolornew.png}
%\caption{Concurrence for single qubit reduced density matrix state vs time at E=1 and N=4}
%\end{figure}
 
%
%\begin{figure}[hbtp]
%\centering
%\includegraphics[scale=0.45]{concurrence1.png}
%\caption{Variations of the concurrence(across system partition of $A_1 | A_2 A_3 ... A_N $) and the probability of success as a function of time at parameter values $E=1$ and $N=4$.}
%\end{figure}

\begin{figure}[hbtp]
\centering
\includegraphics[scale=0.5]{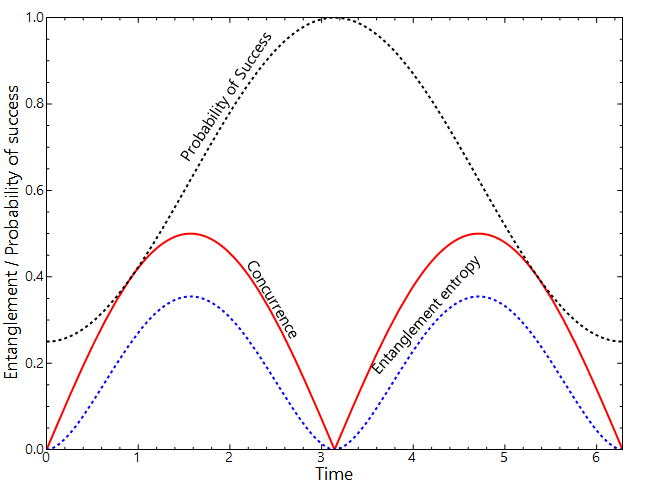}
\caption{Variations of the entanglement entropy, the concurrence (across system partition of $A_1 | A_2 A_3 ... A_N $) and the probability of success as a function of time at parameter values $E=1$ and $N=4$.}
\label{fig:entanglement}
\end{figure}

The entanglement entropy and concurrence across the system partition of $A_1 | A_2 A_3 ... A_N $ are calculated analytically and their variation with time along with the probability of success is shown in Fig. \ref{fig:entanglement}. Note that both entanglement measures go to zero as the success probability peaks to 1.

Rate of change of the concurrence with time is found to be
\begin{equation}
\frac{dC(\rho_{A_1| A_2 A_3 ... A_N })}{dt} = \frac{E\sqrt{2(N-2)}}{N} \cos{\left(\frac{2 E t}{\sqrt{N}}\right)}.
\end{equation}

One can see that 
\begin{equation}
\frac{dC(\rho_{A_1| A_2 A_3 ... A_N })}{dt} \approx E \sqrt{\frac{2}{N}},
\end{equation}
by neglecting O($\frac{1}{N}$) terms for N $\gg$ 1. Therefore, for a large database size (N $\gg$ 1), the rate of concurrence goes to zero, hence implying that a very small amount of entanglement is generated during the search process.\\

\begin{figure}
\centering     %%% not \center
\subfigure[Figure a]{\includegraphics[width=80mm]{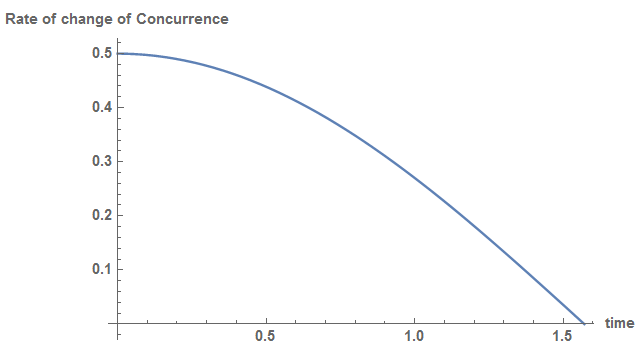}}
\subfigure[Figure b]{\includegraphics[width=80mm]{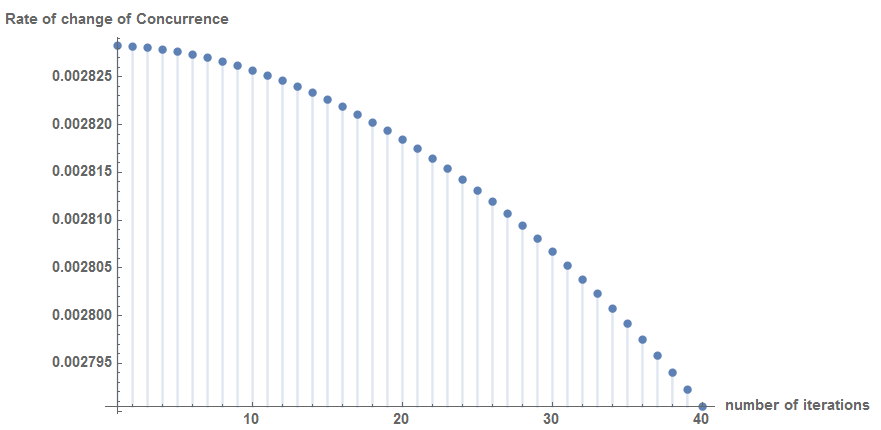}}
\caption{Rate of change of concurrence with time in (a) the analog analogue of Grover search and (b) the original Grover search, across system partition of $A_1 | A_2 A_3 ... A_N $.}
\label{fig:dcbydt}
\end{figure}

Fig. \ref{fig:dcbydt} shows the analogous rate of change between the discrete and the continuous Grover search and suggests that the mapping chosen in subsection \ref{sec:mapping} preserves the properties of the original Grover search algorithm.\\

%
%\section{Concurrence for any two qubit reduced density matrix}
%\label{sec:2qubitconcurrence}

\subsection{Monogamy score}
\label{sec:monogamy}
The amount of quantum correlations that can be shared amongst the subsystems of a multipartite quantum state is captured by the idea of monogamy. The central idea of monogamy is that entanglement cannot be freely shared. Precisely put, \textit{if two qubits A and B are maximally entangled then they cannot be entangled at all with a third qubit C} \cite{ref37}. 

For the tripartite case, if $\cal{C}$ is a bipartite quantum correlation measure, then this  measure is said to be monogamous (or satisfy monogamy) for a tripartite quantum state \(\rho_{ABC}\), if the following condition holds
\begin{eqnarray}
\mathcal{C}(\rho_{A|BC}) \geq \mathcal{C}(\rho_{AB}) +  \mathcal{C}(\rho_{AC}).  
\end{eqnarray}
Here $\mathcal{C}(\rho_{AB})$ is the quantum correlation (with respect to the correlation measure \(\mathcal{C}\)) between subsystems $A$ and $B$, $\mathcal{C}(\rho_{AC})$ is the quantum correlation between subsystems $A$ and $C$, and $\mathcal{C}(\rho_{A:BC})$ is 
quantum correlation between subsystem $A$ and subsystems $B$ and $C$ taken together. For example, if we have a three-qubit pure state $\rho_{ABC} = (|\Psi \rangle \langle \Psi|)_{ABC}$ and a quantum correlation measure say concurrence then it is known that 
\begin{equation}
\mathcal{C}^2_{AB} + \mathcal{C}^2_{AC} \leq 4 \text{ det } \rho_A.
\end{equation}
Even though $BC$ is a two qubit subsystem with a four dimensional subspace, it can be shown that the support of $\rho_{BC}$ is spanned by the eigenvectors corresponding to \textit{at most} two non-zero eigenvalues of the reduced density matrix $\rho_{BC}$, and hence it effectively becomes a two-dimensional space. This allows one to treat the bipartition of $A$ and $BC$ as an effective two-qubit system whose concurrence, $\mathcal{C}_{A|BC}$, is simply given by $2 \sqrt{\text{det } \rho_A}$. Substituting the value of $\mathcal{C}_{A|BC}$ above, we get the Coffman, Kundu, Wootters (CKW) inequality \cite{ref48}
\begin{equation}
\mathcal{C}^2_{AB} + \mathcal{C}^2_{AC} \leq \mathcal{C}^2_{A|BC},
\end{equation}
which suggests that the concurrence is a monogamous entanglement measure.\\

This also leads to the concept of quantum monogamy score, which, for a given bipartite quantum correlation measure, is defined as 
\begin{equation}
 \delta \mathcal{C} \equiv \mathcal{C}(\rho_{A:BC}) - \mathcal{C}(\rho_{AB}) -  \mathcal{C}(\rho_{AC}) \geq \text{0.}
\end{equation}
If a tripartite state satisfies entanglement monogamy then its monogamy score is positive or else it is negative. \\

Another measure of bipartite entanglement is the entanglement of formation (EoF)\cite{ref39}, which is closely related to two-qubit concurrence. Consider a bipartite quantum state $\rho_{AB}$, and the  ensemble $\{p_i,|\psi_i\rangle\}$ denoting a possible pure state decomposition of $\rho_{AB}$, satisfying $\rho_{AB}=\sum_{i}p_i|\psi_i\rangle\langle\psi_i|$. The EoF is defined as 
\begin{eqnarray}
E_f(\rho_{AB})=\underset{\{p_i,|\psi_i\rangle\}}{\min} \sum_{i} p_i S(\text{Tr}_B(|\psi_i\rangle\langle\psi_i|)),
\end{eqnarray}
where $S(\text{Tr}_B(|\psi_i\rangle\langle\psi_i|))$ is the von Neumann entropy of the reduced density matrix corresponding to the $A$ subsystem of $\rho_{AB}$. For a two-qubit mixed state $\rho_{AB}$, $E_f(\rho_{AB})=H \left((1+\sqrt{1-\mathcal{C}_{AB}^2})/2\right)$, where $H(x)=-x \log_2 x-(1-x)\log_2(1-x)$ is the binary entropy function. The EoF, being a concave function of squared concurrence, does not obey the CKW inequality. However, the square of the EoF does obey the same relation as the squared concurrence for tripartite systems \cite{ref49}.

\subsection{Monogamy in analog Grover search}
The discrete analogue of analog Grover search algorithm satisfies the squared concurrence monogamy calculated as
\begin{equation}
\begin{split}
C^2 (\rho_{A_1|A_2 A_3 ... A_N}) - C^2(\rho_{A_1 A_2}) - C^2(\rho_{A_1 A_3})  \\  - ... - C^2(\rho_{A_1 A_N}) \geq 0. \end{split}
\end{equation}

The reduced density matrix for two qubits is calculated by tracing out the $(n-2)$ qubits 
\begin{widetext}
\begin{equation}
\rho_{AB} = 
\begin{bmatrix}
 \frac{\left(\frac{N}{4}-1\right) \beta^2}{N-1}+ \abs{\alpha}{}^2 & \frac{\left(\frac{N}{4}-1\right) \beta^2}{N-1}+\frac{\alpha \beta}{\sqrt{N-1}} & \frac{\left(\frac{N}{4}-1\right) \beta^2}{N-1}+\frac{\alpha \beta}{\sqrt{N-1}} & \frac{\left(\frac{N}{4}-1\right) \beta^2}{N-1}+\frac{\alpha \beta}{\sqrt{N-1}} \\
 \frac{\left(\frac{N}{4}-1\right) \beta^2}{N-1}+\frac{\alpha^{*} \beta}{\sqrt{N-1}} & \frac{\beta^2 N}{4 (N-1)} & \frac{\beta^2 N}{4 (N-1)} & \frac{\beta^2 N}{4 (N-1)} \\
 \frac{\left(\frac{N}{4}-1\right) \beta^2}{N-1}+\frac{\alpha^{*} \beta}{\sqrt{N-1}} & \frac{\beta^2 N}{4 (N-1)} & \frac{\beta^2 N}{4 (N-1)} & \frac{\beta^2 N}{4 (N-1)} \\
 \frac{\left(\frac{N}{4}-1\right) \beta^2}{N-1}+\frac{\alpha^{*} \beta}{\sqrt{N-1}} & \frac{\beta^2 N}{4 (N-1)} & \frac{\beta^2 N}{4 (N-1)} & \frac{\beta^2 N}{4 (N-1)} \\
\end{bmatrix}
\end{equation}
\end{widetext}

The concurrence of an arbitrary two-qubit state is calculated according to the formula in subsection \ref{subsec:concurrence}, using the spin flipped qubit. The eigenvalues for the density matrix $\rho \gamma$ (i.e, the density matrix obtained after multiplying with the spin-flipped qubit for $\rho_{AB}$) are\\ \\
$\lambda_1 = 0$ , $\lambda_2 = 0$ , $\lambda_3 = \frac{1}{4N} \left( \frac{N+4}{4} - \sqrt{N} \right) \sin^2{(\frac{2 E t}{\sqrt{N}})}$ and $\lambda_4 = \frac{1}{4N} \left( \frac{N+4}{4} + \sqrt{N} \right) \sin^2{(\frac{2 E t}{\sqrt{N}})}$.
Therefore, the two qubit concurrence is given by
\begin{small}
\begin{equation}
\mathcal{C}(\rho_{AB}) = \left( \frac{1}{\sqrt{N}} \right) \abs{\sin{(\frac{2 E t}{\sqrt{N}})}}.
\label{eqn:twoqubitconc}
\end{equation}
\end{small}

\begin{figure}[hbtp]
\centering
\includegraphics[scale=0.5]{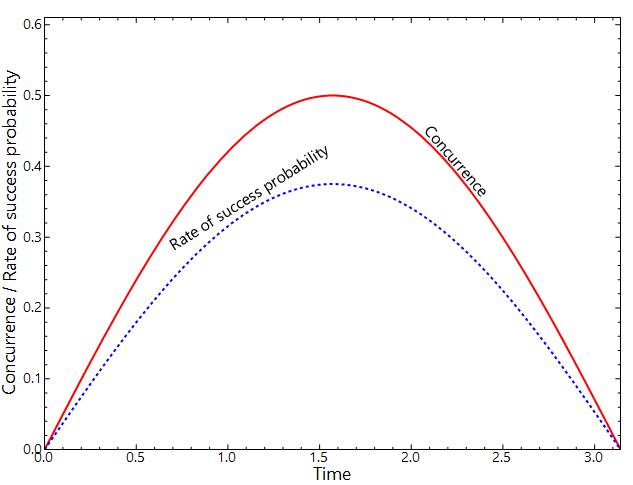}
\caption{Variations of the two-qubit concurrence (i.e., across system partition of $A_1  A_2 | A_3 ... A_N $) and the rate of success probability as a function of time at parameter values $E=1$ and $N=4$.}
\label{fig:concurrenceandrate}
\end{figure}

This is the pairwise entanglement in the analog Grover search. The evolution of the pairwise entanglement is calculated numerically and the result is shown in Fig. \ref{fig:concurrenceandrate} along with the rate of success probability in the search algorithm. The two peak simultaneously suggesting that entanglement is indeed necessary for the discrete analogue of analog Grover search algorithm. 

For the multipartite system, in particular, the pairwise entanglement sharing and other pairwise correlations are monogamous; when $n$ tends to infinity all of the pairwise entanglement vanishes as seen from Eq. (\ref{eqn:twoqubitconc}).

Since in the analog Grover search case, the concurrence between any two pair of qubits is the same, i.e., $\mathcal{C}(\rho_{ij}) = \mathcal{C}(\rho_{kl}) \text{ } \forall \text{ pairs of qubits } ij, kl$; as a result, the monogamy score between the $n$ qubits reduces to the following
\begin{equation}
\delta \mathcal{C} \equiv \mathcal{C} \left(\rho_{A:BC...}\right) - (n-1) \text{ }\mathcal{C}(\rho_{AB}) .
\end{equation}

The monogamy score is, thus, given by
\begin{equation}
\delta \mathcal{C} = \left( \frac{N-2}{2N} - \frac{1}{N} \log_2{\frac{N}{2}} \right) \sin^2\left(\frac{2 E t}{\sqrt{N}}\right)
\end{equation}

\begin{figure}[hbtp]
\centering
\includegraphics[scale=0.5]{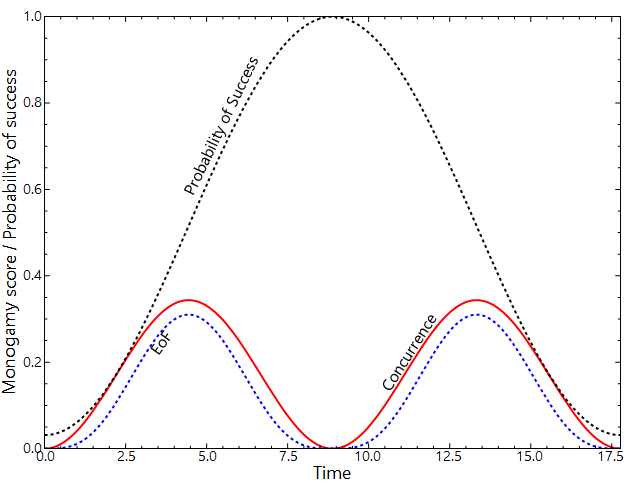}
\caption{Variations of the monogamy score for the concurrence, the monogamy score for the squared entanglement of formation and the probability of success as a function of time at parameter values $E=1$ and $N$ = 32.}
\end{figure}
Similarly, the squared entanglement of formation satisfies a monogamy inequality. We do not provide the expression for this as it is simply too long. Monogamy inequality bounds the amount of pairwise entanglement that can be shared between multiple qubits, and we can see that the discrete analog of Grover search algorithm satisfies two monogamy inequalities for arbitrary times. \\

\section{Conclusion}
\label{sec:conclusion}
To summarize, in this paper we have explored the role of quantum coherence and monogamy of entanglement in the discrete analogue of the analog analogue of Grover search algorithm. Using the $l_1$-norm and the relative entropy of coherence, it was shown that coherence acts as a signature for the success of the analog Grover search algorithm. It was also shown that a maximally coherent state ends up into an incoherent state as the search algorithm evolves and hence $n$-maximal coherence is actually consumed during the search process. 

The variation of entanglement was also quantified and the analogous rate of change of concurrence between the discrete and analog Grover search algorithms suggests that our mapping preserves the original behavior of the algorithm. The pairwise entanglement was shown to peak simultaneously with the rate of success probability as evidence that entanglement is indeed necessary, for the pure state implementation of analog Grover search algorithm.  The pairwise entanglement also suggested a monogamous behavior of quantum correlations in the analog Grover search and it is then shown that the discrete analogue of analog Grover search satisfies the entanglement monogamy inequality for both entanglement measures namely the concurrence and the squared entanglement of formation, for all times during the search algorithm. \\

Note: After the completion of this work, the authors noticed the paper \cite{similar}, where similar results about coherence have been obtained independently by Hai-Long Shi, Si-Yuan Liu, Xiao-Hui Wang, Wen-Li Yang, Zhan-Ying Yang and Heng Fan.
 
\section{Acknowledgments}
Namit Anand would like to acknowledge the hospitality of the Harish-Chandra Research Institute for allowing him to use their facilities during several visits made as a summer student over the last year during the preparation of this manuscript.

\end{document}